\documentclass[english,notitlepage,nofootinbib,superscriptaddress]{revtex4-1}
\usepackage[T1]{fontenc}
\usepackage[latin9]{inputenc}
\usepackage{color}
\usepackage{amsmath}
\usepackage{amssymb}
\usepackage{graphicx}
\usepackage{esint}

\def\gs{\mathrel{
   \rlap{\raise 0.511ex \hbox{$>$}}{\lower 0.511ex \hbox{$\sim$}}}}
\def\ls{\mathrel{
   \rlap{\raise 0.511ex \hbox{$<$}}{\lower 0.511ex \hbox{$\sim$}}}}

\makeatletter

\providecommand{\tabularnewline}{\\}

\usepackage[percent]{overpic}

\makeatother

\usepackage{babel}
\begin{document}

\title{Probing neutrino coupling to a light scalar with coherent neutrino scattering}

\date{\today}

\author{Yasaman Farzan}
\affiliation{Institute for Research in Fundamental Sciences (IPM),
P.O. Box 19395-5531, Tehran, Iran}
\author{Manfred Lindner}
\affiliation{Max-Planck-Institut f\"ur Kernphysik, Postfach
103980, D-69029 Heidelberg, Germany}
\author{Werner Rodejohann}
\affiliation{Max-Planck-Institut f\"ur Kernphysik, Postfach
103980, D-69029 Heidelberg, Germany}
\author{Xun-Jie Xu}
\affiliation{Max-Planck-Institut f\"ur Kernphysik, Postfach
103980, D-69029 Heidelberg, Germany}

\begin{abstract}
\noindent
Large neutrino event numbers in future experiments measuring
coherent elastic neutrino nucleus scattering allow precision
measurements of standard and new physics. We analyze the current and
prospective limits
of a light scalar particle coupling to neutrinos and quarks,
using COHERENT and CONUS as examples. Both lepton number conserving
and violating interactions are considered.
It is shown that current (future) experiments can probe for scalar masses of a few MeV
couplings down to the level of $10^{-4}$ $(10^{-6})$. Scalars with masses around
the neutrino energy allow to determine their mass via a characteristic  
spectrum shape distortion.
Our present and future limits are compared with constraints from supernova evolution,
Big Bang nucleosynthesis and neutrinoless double beta decay.
We also outline  UV-complete underlying models that include a
light scalar with coupling to quarks for both
lepton number violating and conserving coupling to neutrinos.

\end{abstract}
\maketitle

\section{Introduction}

The observation of Coherent Elastic neutrino-Nucleus Scattering (CE$\nu N$S) by the COHERENT experiment \cite{Akimov:2017ade} has opened a new
window to probe Standard Model and beyond the Standard Model
physics. Those include the determination of the Weinberg angle at low
energies \cite{Scholberg:2005qs}, nuclear physics parameters \cite{Cadeddu:2017etk},
searches for a magnetic moment of the neutrino \cite{Dodd:1991ni,Kosmas:2015sqa}, for light sterile
neutrinos \cite{Anderson:2012pn,Dutta:2015nlo,Canas:2017umu,Kosmas:2017zbh}, or for neutrino exotic interactions, be it the vector type
\cite{Barranco:2005yy,Scholberg:2005qs,Barranco:2007tz,
Dutta:2015vwa,Lindner:2016wff,Shoemaker:2017lzs,Liao:2017uzy,
Coloma:2017ncl,Dent:2017mpr} or of
other Lorentz-invariant types \cite{Lindner:2016wff,Kosmas:2017tsq}.

In the present paper we focus on new neutrino physics caused by a new light
scalar. Such a
particle can participate in coherent neutrino-nucleus scattering, and
interestingly modify the nuclear recoil spectrum in a characteristic
manner, both for a light scalar as well as heavy one. Most of our
study will focus on the light scalar case. Light new physics is of
course motivated by the lack of signals in collider
experiments, and its consequences in coherent neutrino-nucleus scattering have been
mentioned before
\cite{Dent:2016wcr,Cui:2017ytb,Shoemaker:2017lzs,Liao:2017uzy,Coloma:2017ncl,AristizabalSierra:2017joc,Ge:2017mcq}. Mostly 
it was used that light physics does not suffer from limits on
neutrino non-standard interactions from high energy scattering
experiments such as CHARM-II, as stressed in
\cite{Farzan:2015doa,Farzan:2016wym}. Discovery limits on light
particles in CE$\nu N$S have been discussed in Refs.\
\cite{Dent:2016wcr,Shoemaker:2017lzs,Ge:2017mcq}, indicating already before the observation by COHERENT that CE$\nu N$S could provide strong constraints on light mediators.
In our work, as mentioned before, we focus on a new scalar particle
and obtain the current limits from COHERENT on its mass and coupling
with SM particles. We also consider explicit future realizations of
this experiment and also of CONUS, which will use reactor
antineutrinos to probe coherent scattering. The characteristic
distortion of the spectrum shape for light scalars with masses around
the neutrino energy allows to reconstruct the scalar mass, which we
explicitly demonstrate. We also compare our limits with existing ones
from a variety of sources in particle and astroparticle
physics. UV-complete models that may be behind the existence of such
light scalar particles are also outlined.

This paper is organized as follows: In Sec.\ \ref{sec:model}, we present the framework of
a light scalar with couplings to neutrinos and quarks, discuss the
coherent scattering cross section and include a discussion
of form factors when the coupling to nucleons is considered. In Sec.\ \ref{sec:limits},  we discuss the constraints on such
scalar particles
from various particle and astroparticle physics observables. In Secs.\
\ref{sec:Constraints} and \ref{sec:disc} we discuss and interpret the implications of various possible observations. In
Sec.\ \ref{sec:concl}, we summarize our findings.
The calculational details and outlines for  UV-complete gauge
invariant models are presented in Appendices.


\section{Light scalar interactions in coherent neutrino-nucleus scattering \label{sec:model}}

In this section, we first introduce the possible interactions of neutrinos with quarks (or nuclei) mediated by a light scalar boson and then discuss the corresponding cross section of coherent neutrino-nucleus scattering.

\subsection{New scalar interactions}
We consider a scalar field denoted by $\phi$ which couples to neutrinos. There are two possibilities for such coupling, namely lepton number violating (LNV) and conserving (LNC) couplings. The latter possibility requires the
presence of right-handed neutrinos:
\begin{equation}
{\cal L}_{\rm LNC} \equiv
\bar \nu \left( C_\nu + i D_\nu \gamma_5 \right)\nu \phi  =
y_\nu \phi \bar{\nu}_R\nu_L +{\rm H.c.},  \label{LNC-Eq}
\end{equation}
where $y_{\nu}=C_{\nu}-iD_{\nu}$. For simplicity, throughout this
paper  we implicitly assume that $\phi$ is a real field but most discussion and overall behavior of the bounds and limits remains valid for a complex $\phi$ as well. For real $\phi$, the hermicity of the Lagrangian implies that $C_\nu$ and $D_\nu$ are real.
Note that the couplings
have flavor indices, suppressed here for clarity.
The lepton number violating form of the interaction can be written in
analogy as
\begin{equation}
{\cal L}_{\rm LNV} \equiv
\frac{y_\nu}{2} \phi \bar{\nu}_L^c \nu_L +{\rm H.c} =\frac{y_\nu}{2}
\phi \nu_L^T C\nu_L +{\rm H.c.} \label{LNV-Eq}
\end{equation}
In both Lagrangians $y_\nu$ can in general be a complex number.\\

The same scalar field can couple to quarks. Since in coherent neutrino
scattering we are concerned with the effective coupling of $\phi$ with
the whole nucleus $N$ we write the Lagrangian as:
\begin{equation}
{\cal L}_{N\phi}\equiv\overline{\psi_{N}}
\Gamma_{N\phi}\psi_{N}\phi,\label{eq:ss}
\end{equation}
where $\psi_{N}$ is the Dirac spinor of the nucleus, assuming it is a spin-1/2 particle\footnote{The actual spin of the nucleus
can take other values but the difference of the cross section
is suppressed by $E_{\nu}^{2}/M^{2}$---see
the appendix in \cite{Lindner:2016wff}. }. We can write
\begin{equation}
\Gamma_{N\phi}\equiv C_{N}+D_{N}i\gamma^{5}.\label{eq:s-2}
\end{equation}
Again, for real $\phi$ hermicity of the Lagrangian demands $C_{N}$ and $D_{N}$ to be real numbers.
The conversion from  fundamental quark couplings ($C_{q}$, $D_{q}$) to the
effective coupling  ($C_{N}$, $D_{N}$) will be discussed later. Note
that we here consider both scalar and pseudo-scalar interactions. In
what follows, the latter contribution is usually very much suppressed,
and essentially only the scalar contribution is what matters.

In summary, the Lagrangian (in addition to the SM) responsible for coherent neutrino scattering is
\begin{equation}
{\cal L}\supset{\cal L}_{\nu\phi}+{\cal L}_{N\phi}
-\frac{1}{2}m_{\phi}^{2}\phi^{2}
-M\overline{\psi_{N}}\psi_{N},\label{eq:s}
\end{equation}
where ${\cal L}_{\nu\phi}$ can be either Eq.~(\ref{LNC-Eq}) or Eq.~(\ref{LNV-Eq}).   The masses of the scalar and nucleus are respectively denoted by  $m_{\phi}$ and $M$.
In App.\ \ref{model-EW}, we present examples in which
couplings to neutrinos and quarks can be embedded in electroweak symmetric models.

\subsection{Cross Section \label{CS}}
The first thing to notice is that the Yukawa interaction of both LNC
and LNV forms (either Eq.\ (\ref{LNC-Eq}) or Eq.\ (\ref{LNV-Eq}))
leads to chirality-flipping
scattering which will not interfere with chirality-conserving SM weak
interactions\footnote{More generally, as it has been studied in \cite{Rodejohann:2017vup},
there is no interference in neutrino scattering between vector (or
axial-vector) form interactions and other forms of interactions, including
(pseudo-)scalar and tensor.}. Thus, one can separate the cross section
into two parts containing the pure SM and the new physics
contributions:
\begin{equation}
\frac{d\sigma}{dT}=\frac{d\sigma_{{\rm SM}}}{dT}+\frac{d\sigma_{\phi}}{dT},\label{eq:s-7}
\end{equation}
where $T$ denotes the recoil energy.
The SM cross section, assuming full
 coherence is given by \cite{Freedman:1973yd}
\begin{equation}
\frac{d\sigma_{{\rm
      SM}}}{dT}=\frac{G_{F}^{2}M\left[N-(1-4s_{W}^{2})Z\right]^{2}}{4\pi}\left(1-\frac{T}{T_{\max}}\right), 
\mbox{ where }
T_{{\rm max}}(E_{\nu})=\frac{2E_{\nu}^{2}}{M+2E_{\nu}}.\label{eq:s-10}
\end{equation}
The coherent scattering mediated by the light scalar,
independent of whether the new scalar
interaction is of the LNC or LNV form, is (the derivation is given in 
appendix \ref{sec:calc}):
\begin{equation}
\frac{d\sigma_{\phi}}{dT}=\frac{MY^{4}A^{2}}{4\pi(2MT+m_{\phi}^{2})^{2}}\left[\frac{MT}{E_{\nu}^{2}}+{\cal O}\left(\frac{T^{2}}{E_{\nu}^{2}}\right)\right],\label{eq:s-9}
\end{equation}
where we have defined
\begin{equation}
Y^{4}\equiv\frac{C_{N}^{2}}{A^{2}}|y_{\nu}|^2.\label{eq:s-11}
\end{equation}
Here $C_N$ is the coupling of the scalar with the nucleus, whose
connection to the fundamental quark couplings is discussed in the next
subsection.  
The division by the atomic number in the definition of $Y$ makes it almost independent
of the type of  nucleus---cf. Eq.~(\ref{eq:s-37}) and Eq.~(\ref{eq:s-38}).
The cross section has little  dependence\footnote{
The explicit form of the negligible term ${\cal
  O}\left(T^{2} /E_{\nu}^{2}\right)$ is actually $\left(1+D_{N}^{2}/C_{N}^{2}\right)\frac{T^{2}}{2E_{\nu}^{2}}$---see Eq.\ (\ref{eq:s-9-1}) in the appendix.
} on  $D_{N}$ because
the pseudo-scalar contribution is suppressed by the ${\cal
  O}\left(T^{2} /E_{\nu}^{2}\right)$
term. This is in analogy to dark matter direct detection where dark matter nucleon interactions mediated by a pseudo-scalar is well known to be suppressed.


Obviously scalar interactions lead to a  spectral shape different from that
in the SM case, see e.g.\ \cite{Lindner:2016wff}, where an
effective scalar interaction, corresponding to $m_\phi^2 \gg M T$, is
considered.  For a light scalar
under discussion here we see that additional modifications of the
spectrum are possible, in particular if the scalar mass is of the same order or smaller than the typical momentum transfer $M T \sim
E_\nu^2$.


\subsection{From the fundamental couplings to the effective couplings\label{Ff}}

 The effective couplings $C_{N}$ and $D_{N}$ originate from  fundamental couplings of $\phi$ with the quarks. The connection
between the effective couplings and the fundamental couplings has
been well studied in spin-independent dark matter direct detection.
Essentially, one needs to know the scalar form factors of quarks in
the nuclei. We refer to \cite{Belanger:2008sj} for the details and
summarize the relevant results below.

Since the pseudo-scalar coupling $D_{N}$ has no effect on CE$\nu N$S
{[}cf.\ Eq.~(\ref{eq:s-11}){]}, we focus on the scalar
coupling, $C_N$.
Taking the fundamental scalar interaction of quarks with $\phi$
to be of the form
\begin{equation}
{\cal L}\supset\sum_{q}C_{q}\overline{q}q\phi,\label{eq:s-13}
\end{equation}
the effective coupling $C_{N}$ is related to $C_{q}$ by
\begin{equation}
C_{N}=ZC_{p}+(A-Z)C_{n},\label{eq:s-21}
\end{equation}
where the couplings to protons and neutrons are
\begin{equation}
C_{p}=m_{p}\left[\sum_{q}C_{q}\frac{f_{q}^{p}}{m_{q}}\right],\ C_{n}=m_{n}\left[\sum_{q}C_{q}\frac{f_{q}^{n}}{m_{q}}\right].\label{eq:s-39}
\end{equation}
Here $m_{p}=938.3$ MeV and $m_{n}=939.6$ MeV are masses of proton
and neutron; $Z$ and $A-Z$ are proton and neutron numbers in the
nucleus; $m_{q}$ are quark masses; $f_{q}^{p}$ and $f_{q}^{n}$
are the scalar form factors in protons and neutrons. According to
the updated data for the $u$ and $d$ quarks from \cite{Crivellin:2013ipa,Hoferichter:2015dsa}
and the data for the $s$ quark from \cite{Junnarkar:2013ac}, the
form factors are:
\begin{equation}
f_{d}^{p}=0.0411\pm0.0028,\ f_{u}^{p}=0.0208\pm0.0015,\ f_{s}^{p}=0.043\pm0.011,\ f_{c}^{p}\approx f_{b}^{p}\approx f_{t}^{p}\approx\frac{2}{27}(1-f_{d}^{p}-f_{u}^{p}-f_{s}^{p})\approx0.066,\label{eq:s-22}
\end{equation}
\begin{equation}
f_{d}^{n}=0.0451\pm0.0027,\ f_{u}^{n}=0.0189\pm0.0014,\ f_{s}^{n}=0.043\pm0.011,\ f_{c}^{n}\approx f_{b}^{n}\approx f_{t}^{n}\approx\frac{2}{27}(1-f_{d}^{n}-f_{u}^{n}-f_{s}^{n})\approx0.066,\label{eq:s-23}
\end{equation}
where $f_{c,\thinspace b,\thinspace t}$ are approximately the same
for the heavy quarks because their contributions come from heavy
quark loops that couple $\phi$ to the gluons \cite{Shifman:1978zn}. Notice that despite the different
quark content, the  couplings of proton and neutron to the scalar turn out to be almost
equal: $|(C_n-C_p)/C_n|=  {\cal O}(10\%)$.
If $C_{q}$ for the heavy quarks are of the same order of magnitude
as for the light quarks, then the contributions of heavy quarks are negligible
due to suppression by their masses. However, if $C_q \propto m_q$ (as in the case that the
$\phi$ coupling to the quarks comes from the mixing of a scalar singlet with the SM Higgs),
the contribution from all flavors will be comparable. Taking the following quark
masses \cite{Olive:2016xmw}:
\[
m_{d}=4.7\ {\rm MeV},\ m_{u}=2.2{\rm \ MeV},\ m_{s}=96\ {\rm MeV},\ m_{c}=1.27{\rm \ GeV},\ m_{b}=4.18\ {\rm GeV},\ m_{t}=173.2\ {\rm GeV},
\]
we obtain
\begin{equation}
m_{p}\left(\frac{f_{d}^{p}}{m_{d}},\ \frac{f_{u}^{p}}{m_{u}},\ \frac{f_{s}^{p}}{m_{s}},\ \frac{f_{c}^{p}}{m_{c}},\ \frac{f_{b}^{p}}{m_{b}},\ \frac{f_{t}^{p}}{m_{t}}\right)\approx(8.2,\thinspace8.9,\thinspace0.42,\thinspace4.9\times10^{-2},\thinspace1.5\times10^{-2},\thinspace3.6\times10^{-4})\,,\label{eq:s-24}
\end{equation}
\begin{equation}
m_{n}\left(\frac{f_{d}^{n}}{m_{d}},\ \frac{f_{u}^{n}}{m_{u}},\ \frac{f_{s}^{n}}{m_{s}},\ \frac{f_{c}^{n}}{m_{c}},\ \frac{f_{b}^{n}}{m_{b}},\ \frac{f_{t}^{n}}{m_{t}}\right)\approx(9.0,\thinspace8.1,\thinspace0.42,\thinspace4.9\times10^{-2},\thinspace1.5\times10^{-2},3.6\times10^{-4})\,.\label{eq:s-25}
\end{equation}
For Ge and CsI targets, taking the average values of $(Z,\thinspace A)$
as  $(32,\thinspace72.6)$ and $(54,\thinspace130)$,
we can evaluate the explicit dependence of $C_{N}$ on the $C_{q}$:
\begin{equation}
C_{N}=\begin{cases}
10^{2}\times(6.3\,C_{d}+6.1\,C_{u})+(30.5\,C_{s}+3.5\,C_{c}+1.1\,C_{b}+2.5\times10^{-2}\,C_{t}) & ({\rm Ge})\\
10^{3}\times(1.1\,C_{d}+1.1\,C_{u})+(54.7\,C_{s}+6.3\,C_{c}+1.9\,C_{b}+4.7\times10^{-2}\,C_{t}) & ({\rm CsI})
\end{cases}.\label{eq:s-27}
\end{equation}
This means that from the fundamental coupling to the effective coupling an
amplification by a factor of ${\cal O}(10^{2})$ or ${\cal O}(10^{3})$ can be present.
The definition of $Y$ in Eq.~(\ref{eq:s-11}), in terms of the
fundamental couplings, can be rewritten as
\begin{equation} \label{eq:Y}
Y\equiv \sqrt{ 
\frac{|C_N y_\nu|}{A}
}
=\sqrt{\left|\left(\frac{A-Z}{A}\,C_{n}+\frac{Z}{A}\,
C_{p}\right)y_{\nu}\right|}.
\end{equation}
The dependence of $Y$ on the types of targets is weak because for heavy nuclei, $\frac{A-Z}{A}$ and  $\frac{Z}{A}$ are typically close to $1/2$.
For example, taking the average values of $(Z,\thinspace A)$ for Ge and CsI
targets, we get
\begin{equation}
Y_{{\rm Ge}}\approx\sqrt{|(0.56\,C_{n}+0.44\,C_{p})y_{\nu}|}\,,\label{eq:s-37}
\end{equation}
\begin{equation}
Y_{{\rm CsI}}\approx\sqrt{|(0.58\,C_{n}+0.42\,C_{p})y_{\nu}|}\,.\label{eq:s-38}
\end{equation}
When comparing the sensitivities of CE$\nu N$S experiments using different
targets, we will ignore the small difference  and assume $Y_{{\rm Ge}}\approx Y_{{\rm CsI}}$.

\section{\label{sec:limits} Existing Bounds from Particle and Astroparticle Physics}

In this section we review the relevant bounds on $C_n$, $y_\nu$ or on their product ($C_ny_\nu$) from
various observations and experiments other than coherent
scattering.  As we shall see, the bounds on the hadronic couplings are on
$C_n={\cal R}[\Gamma_{n\phi}]$ (not on $D_n$) but the bounds on neutrino couplings are on $|y_\nu|^2=C_\nu^2+D_\nu^2$. The difference originates from the fact that while nuclei in the considered setups are non-relativistic, neutrinos are ultra-relativistic.
\begin{itemize}
\item
\textbf{Bounds on $C_n$ from neutron nucleus scattering:}
In the mass range of our interest, the strongest bounds on $C_n$ come from low energy neutron scattering off nuclei \cite{Kamiya:2015eva,Pokotilovski:2006up,Leeb:1992qf}, in particular using Pb as target.
The effect of a new scalar would be to provide
a Yukawa-type scattering potential whose effect can be constrained.
Notice that like the case of CE$\nu N$S, since in these setups the nuclei are non-relativistic, their dominant sensitivity is only to $C_n$ (not to $D_n$).

\item
\textbf{Bounds on $y_\nu$ from meson decay:} If the new light boson couples to neutrinos, it can open new decay modes for mesons such as $K^+ \to l^+ \nu \phi$ or
$\pi^+ \to l^+ \nu \phi$. The scalar will eventually decay into a
neutrino pair appearing as missing energy. From the absence of a
signal for such decay modes, bounds  of order $10^{-3}$ on
$(\sum_\alpha |y_\nu|_{e \alpha}^2)^{1/2}$ and $(\sum_\alpha |y_\nu|_{\mu \alpha}^2)^{1/2}$ have been found from different modes \cite{Pasquini:2015fjv}. Notice that as long as
the mass of $\phi$ is much smaller than the meson mass the bound is independent of $m_\phi$. Moreover, the bound similarly applies for the LNV and LNC cases.
In the LNC case, sensitivity is to the combination $|y_\nu|^2= C_\nu^2+D_\nu^2$.
\item \textbf{Bounds from double beta decay:}  A LNV coupling of form $\phi \nu_e^T C\nu_e$ can cause
neutrinoless double beta decay \cite{Rodejohann:2011mu}  in the form of $n+n\to p +p + e^-+e^-+\phi$, of course provided that
$\phi$ is lighter than the $Q$ value of the decaying nucleus. From
double beta decay of $^{136}$Xe with a $Q$ value of 2.4 MeV the following bound is found \cite{Gando:2012pj}:
\begin{equation} \label{ddbb}
(y_\nu)_{ee}<10^{-5} .
\end{equation}
A more recent and a bit weaker bound for $m_\phi<2.03$ MeV
comes from $^{76}$Ge double beta decay \cite{Agostini:2015nwa}.

\item \textbf{Supernova bounds and limits:} Light particles coupled to
  neutrinos and neutrons can affect the dynamics of a proto-neutron
  star in several ways. Before discussing the impact of our particular
  scenario on supernovae, let us very briefly
  review the overall structure of the core of a proto-neutron star in
  the first few 10 seconds after explosion when neutrinos are trapped inside the core (i.e., the mean free path of neutrinos is much smaller than the supernova core radius).
 For more details, the reader can consult the textbook \cite{Raffelt:1996wa}.
In the first $\sim 10$ sec after collapse, the core has a radius of $\sim$ few 10 km and
a matter density of $\rho \sim 10^{14} \, {\rm g}\, {\rm cm}^{-3}$ (comparable to nuclear density).
However, because of the high pressure, most of the nucleons are free. As mentioned before,
neutrinos are trapped inside the core and are thermalized with a temperature of $\sim 10$ MeV.
The core can be hypothetically divided into the inner core with a radius of 10-15 km
and the outer core. Within the inner core, the chemical potential of the $\nu_e$ is about
200 MeV which is much larger than their temperature, implying that
$\nu_e$ are degenerate. The chemical potentials of $\nu_\mu$ and $\nu_\tau$ are zero and their temperatures are equal.
Thus, inside the inner core $n_{\bar{\nu}_e}\ll n_{\nu_\mu}=  n_{\bar\nu_\mu}=
n_{\nu_\tau}= n_{\bar\nu_\tau}\ll n_{\nu_e}$, where $n_{\nu_\alpha}$ is the number density of
$\nu_\alpha$.
In the outer core, the chemical potential of $\nu_e$ decreases and the number densities of
$\nu_e$ and $\bar{\nu}_e$ are almost equal. Notice that the average
energies of $\nu_\mu$,  ${\nu}_\tau$  and their antiparticles throughout the
core are given by their temperature which is of order of few 10 MeV.
The energies of $\nu_e$ and $\bar{\nu}_e$ in the outer core, where the chemical potential vanishes,
are also of the same order but the energy of $\nu_e$ in the inner core is of
order of the chemical potential, 200 MeV. The neutrinos scatter
off nucleons with a cross section of $\sigma\sim G_F^2 E_\nu^2/(4\pi)$.
Considering the high density of nucleons, the mean free path for
neutrinos will be of order of $\lambda\simeq 300$ cm, which is much
smaller than the core radius $R \sim $ few 10 km. The diffusion time, given by
$R (R/\lambda)$, is therefore of order of ${\cal O}(1~{\rm sec})$
to ${\cal O}(10~{\rm sec})$. Neutrinos diffusing out of the
core carry out the binding energy of the star which is of order of $10^{53}$ erg.

Let us now see how  new light particles coupled to neutrinos and matter fields can affect this
picture. First, let us discuss the impact on supernova cooling. If the interaction of the new particle is
very feeble, it cannot be trapped. Thus, if it is produced inside the core
it can exit without hinderance and take energy out of the core leaving no energy for neutrinos to
show up as the observed events of SN1987a.
This sets an upper bound on the coupling of the new particles. On the other hand, if the
coupling is large enough to trap the new particles, the impact on cooling will not be dramatic.
Still, if the new particles are stable, they can diffuse out and, along with neutrinos,
can contribute to supernova cooling. Considering however the theoretical uncertainty in the
evaluation of total binding energy and the observational uncertainty on the energy carried away,
such contributions can be tolerated.
 Thus, supernova cooling consideration, within present uncertainties, can only rule out a
range of coupling between an upper bound and a lower bound.
New interactions can also affect the mean free path of neutrinos $\lambda$ and therefore
the diffusion time $R^2/\lambda$, which roughly speaking coincides with the observable
duration of neutrino emission from a supernova. This sets another limit.
Finally, if there is  a new process that can remove $\nu_e$ and/or convert it to any of
$\bar\nu_e$,  ${\nu}_{\mu(\tau)}$ or $\bar{\nu}_{\mu(\tau)}$, it can have profound effect on
the Equation  of State (EoS) in the inner core. For example, if $\nu_e$ (with energy of 200 MeV) are
converted to $\nu_\mu$, the temperature of $\nu_\mu$ will increase dramatically.
Conversion of $\nu_e$ to $\bar{\nu}_e$ will lead to the production of $e^+$ which annihilates
with electrons inside the core.

  In our case the $\phi$ particles can decay into a
  neutrino pair. The decay length is evaluated to be  approximately
$  (10^{-5}/|y_\nu|)^2(E_\phi/ 10 ~{\rm MeV})(5~{\rm MeV}/m_\phi)^2$ cm,
which is much smaller than the radius of the proto-neutron star ($\sim
$ few 10 km).
The consequences of new interactions on supernova explosions can be
categorized into three effects:
(1)  change of equation of state in case of LNV interaction;
(2) new cooling modes because of right-handed (anti)neutrino emission
in case of LNC interaction;
(3) prolonging the duration of neutrino emission
($R^2/\lambda$) because of a shorter mean free path $\lambda$.
In all these three effects, the neutrino scattering plays a key role.
Neglecting flavor indices, we find
that the neutrino-neutrino scattering cross section is comparable to
the $\nu$-nucleon  scattering cross section if $y_{\nu}  \sim C_n$. The number density of
nucleons is larger than that of neutrinos, thus the $\nu$-nucleon
scattering should be more important. The cross section of the
scattering  due to $\phi$ both for LNV and LNC cases can be estimated as
\begin{equation}\sigma( \stackrel{(-)}{\nu} +n \to \stackrel{(-)}{\nu}
  + n)= \frac{|y_{\nu}| ^2 C_n^2}{   16\pi  E_\nu^2}\left[\log
    \frac{4E_\nu^2+m_\phi^2}{m_\phi^2} -\frac{4
      E_\nu^2}{4E_\nu^2+m_\phi^2}\right].\label{cross-scat}
\end{equation}
The scattering cross section for sterile (anti)neutrinos is given by
the same formula.

Let us discuss the three effects mentioned above one by one.
If $\nu_e$ has a LNV coupling to $\phi$ of the form $(y_{\nu} )_{e
  \alpha} \phi \nu_e^T C\nu_\alpha$, where $\alpha \in \{ e, \mu ,
\tau \}$, $\nu_e$ can be converted into $\bar{\nu}_\alpha$. The
produced antineutrino will be trapped. However, if
\begin{equation} \label{lim} \sigma (\nu_e+n\to \bar{\nu}_\alpha +n)
  \times \frac{\rho}{m_n} \times (10 ~{\rm sec})\gs 1,
\end{equation}
a significant fraction of degenerate $\nu_e$ in the inner core will convert into
antineutrinos, drastically changing the equation of state.
That is, for
\begin{equation} \sqrt{(y_{\nu} )_{e\alpha} C_n}>2\times 10^{-7}
  \sqrt[4]{\frac{E_\nu^2}{\log\frac{4E_\nu^2+m_\phi^2}{m_\phi^2}-\frac{4E_\nu^2}{4E_\nu^2+m_\phi^2}}}\,,
\end{equation}
 in which $E_{\nu_e}\! \sim 200$ MeV, the equation of state of the
 supernova core has to be reconsidered.
In subsequent plots that summarize the limits on our scenario, we call
the associated limit to avoid this feature as ``SN core EoS''.
Since the temperatures of
 $\nu_\mu$ and $\nu_\tau$ are expected to be the same and their
 chemical potentials to be zero, conversion of $\nu_\mu$ and
 $\nu_\tau$ to $\bar{\nu}_\mu$ and $\bar{\nu}_\tau$ due to non-zero
 $(y_{\nu} )_{\mu \mu}$, $(y_{\nu} )_{\mu \tau}$ and $(y_{\nu} )_{\tau \tau}$
 will not change the equation of state.

When the interaction is LNC, scattering will
convert left-handed neutrinos (right-handed antineutrinos) into
sterile right-handed neutrinos (left-handed antineutrinos), which do
not participate in weak interactions. If $\sigma (\rho/m_n)  (10\,{\rm
  sec}) \gs 1$, a significant fraction of the active (anti)neutrinos
will convert into sterile ones. Avoiding this generates an upper limit on the coupling.
If $\sigma (\rho/m_n) R \gs 100$,
the produced sterile neutrinos will be trapped.
The values of $\sqrt{y_{\nu}  C_n}$ between these two limits are
therefore excluded by supernova cooling considerations.
The limits are denoted in Fig.\ \ref{fig:LNC} as ``SN energy loss'' and ``SN $\nu_R$ trapping'', respectively.

Drawing these figures we have
assumed a nominal temperature of 30~MeV for neutrinos
which is the typical energy for all neutrinos in the outer core.
        In case that the neutrino-nucleon scattering cross section due to $\phi$
        exchange becomes comparable to the standard weak cross section
$G_F^2 E_\nu^2/(4\pi)$, the diffusion time $R^2/\lambda$ will be  significantly affected.
This limit is shown in Figs.\ \ref{fig:LNC} and \ref{fig:LNV} as ``SN
$\nu$ diffusion'', and holds for both the LNC and LNV cases.

        \item\textbf{BBN and CMB bounds:} The
          contribution  from the LNC and LNV cases to the additional number of relativistic degrees of
          freedom ($\delta N_{\rm eff}$)  will
          be quite different so in the following, we address them separately.

i) LNV case: In this case, no $\nu_R$ exist so we should only check
for the $\phi$ production. For $m_\phi\gs 1$ MeV, $\phi$ will be
produced at high temperatures but it will decay before neutrino
decoupling without affecting $N_{\rm eff}$ at  the BBN or the
CMB era.
Thus, within the present uncertainties, there is no bound from BBN for $m_\phi\gs 1$ MeV.
Lighter $\phi$, in turn, can contribute to $N_{\rm eff}$ as one scalar degree of
freedom if they enter thermal equilibrium. Taking $n_\nu \sigma(\nu \nu
\to \phi) H^{-1}|_{T=1\,\rm MeV} <1$, we find
\begin{equation}
|y_{\nu}| <5\times 10^{-9} \frac{m_\phi}{\rm MeV}. \label{BBNbo}
\end{equation}
Notice that for $0.1~{\rm MeV}<m_\phi<1~{\rm MeV}$, although $\phi$
decays away before the onset of BBN it still contributes to $N_{\rm eff}$
by warming up the $\nu$ and $\bar{\nu}$ distributions.
At $m_\phi= 1$ MeV, combining the above bound on $y_{\nu} $ from BBN with that from $n$-nucleus scattering,
yields $Y<5 \times 10^{-7}$.  For $m_\phi>1$ MeV, this bound does not apply because $\phi$ decays
into neutrinos before neutrino decoupling from the plasma.
That is why the bound denoted ``BBN + n scat.'' appears as a vertical line in Fig.\ \ref{fig:LNV}.
Our simplified
analysis seems to be in excellent agreement with the results
of \cite{Huang:2017egl} which solves the full Boltzmann equations.

ii) LNC case: In this case, the production of scalars via $\nu \nu \to
\phi$ is not possible. Processes like $\nu \bar{\nu} \to \nu \bar{\nu}
\phi$ or  $\nu N \to \nu N \phi$  can take place but are
suppressed. The $t$-channel process $\nu_L \bar{\nu}_L \to \nu_R
\bar{\nu}_R$ can lead to $\nu_R$ and $\bar{\nu}_R$ production. The
$\nu_L \nu_L \to \nu_R \nu_R$ process can also take place but because of
cancelation between $t$ and $u$ channel diagrams it has a smaller
cross section.  The cross section  of the dominant production mode is
\begin{equation}
\sigma( \nu_L \bar{\nu}_L \to \nu_R \bar{\nu}_R)= \frac{|y_{\nu}| ^4}{32\pi
  m_\phi^2}\frac{1}{x^2} \left(
  \frac{2x(1+x)}{1+2x}-\log(1+2x)\right),
\end{equation}
where $x=2E_\nu^2/m_\phi^2$, in which $E_\nu$ is the energy of
colliding neutrinos in the center-of-mass frame. Taking $n_\nu \sigma
H^{-1} |_{T=3\,{\rm MeV}}\ls0.3$ (3 MeV is the temperature at  neutrino
decoupling and 0.3 is the bound from CMB on $\delta N_{\rm eff}$ \cite{Ade:2015xua}), we
find
\begin{equation}
y_{\nu} <\left. 1.7\times 10^{-5} \left(\frac{m_{\phi}^2}{T}
  \frac{x^2(1+2x)}{2x(1+x)-(1+2x)\log (1+2x)}\right)^{1/4}\right|_{T=3\,{\rm
    MeV}}    ,
\end{equation}
in which $m_\phi$ and $T$ are in MeV and $x=2 T^2/m_\phi^2$. Combining
this bound with the one on $C_n$ 
from neutron-Pb scattering gives the limit denoted ``BBN + n scat.'' in Fig.\ \ref{fig:LNC}.

\end{itemize}

Let us now discuss the effects of forward scattering off nuclei due to the new interaction on neutrino propagation in matter composed of nuclei, $N$. The induced effective mass will be
of chirality flipping form as follows
$$\frac{C_N y_{\nu} }{m_\phi^2} \frac{\rho}{m_N} \left( \nu^TC\nu \ \ \ {\rm or} \ \ \ \bar{\nu}_R\nu_L \right).$$
Notice that   $m_N \simeq A  m_p \simeq Am_n$ and $C_N/m_N \simeq
C_p/m_p$ so we can write the effective mass as $(C_p y_{\nu} /{m_\phi^2}
)({\rho}/{m_p})$. Taking for example the density of the Sun as
$\rho=150\,{\rm g}\, {\rm cm^{-3}}$, we find a shift in the mass of
neutrinos of order of $3 \times 10^{-12} \,{\rm eV}\, (y_{\nu} /10^{-5})
(C_p/10^{-5}) (5~{\rm MeV}/m_\phi)^2$,
which is completely negligible compared to $\Delta m^2/m_\nu$.  One
may wonder why forward scattering due to a possible new gauge boson
with similar mass and coupling has such a large impact on neutrino
propagation in matter, while the present case of a scalar does
not. The reason lies in the different Lorentz structure of the induced
operators.  The vectorial interaction induces a contribution of form
$\bar{\nu}^T \gamma^0 \nu=\nu^\dagger \nu$ which has to be compared
with $m_\nu^2/E_\nu$. In the case of scalar, the matter effects have
the operatorial form as the mass themselves and should be compared to
the mass splitting, $\Delta m^2/m_\nu$.
More detailed discussion can be found in \cite{Bergmann:1999rz}.

\section{Constraints and future Sensitivities from CE$\nu$NS\label{sec:Constraints}}

To collect large statistics at coherent scattering energies, CE$\nu N$S
experiments require intensive and low-energy ($\lesssim50$ MeV) neutrino
fluxes.  Two types of neutrino sources can be invoked to carry out
 CE$\nu N$S experiments: reactor neutrinos ($E_\nu \lesssim8$ MeV) and
pion decay at rest ($E_\nu \lesssim50$ MeV). Two on-going experiments,
CONUS\footnote{https://indico.cern.ch/event/606690/contributions/2591545/attachments/1499330/2336272/Taup2017\_CONUS\_talk\_JHakenmueller.pdf}
and COHERENT \cite{Akimov:2017ade}, adopt these two sources respectively.
In this section, we study the sensitivities of the two experiments
on light scalar bosons.

\subsection{CONUS}
The CONUS experiment uses a very low threshold Germanium detector
setting 17 m away near a nuclear power plant (3.9 GW thermal power)
in Brokdorf, Germany. The total antineutrino flux is
$2.5\times10^{13}\, {\rm s}^{-1}\, {\rm cm}^{-2}$.
Data collection started in 2017 and first results are expected soon.
To study the sensitivity of CONUS, we compute
the event numbers given by
\begin{equation}
N_{i}=\Delta tN_{{\rm Ge}}\int_{T_{i}}^{T_{i}+\Delta T}dT\int_{0}^{8\thinspace{\rm MeV}}dE_{\nu}\Phi(E_{\nu})\theta\left(T_{{\rm max}}(E_{\nu})-T\right)\frac{d\sigma}{dT}\left(T,\thinspace E_{\nu}\right).\label{eq:s-28}
\end{equation}
Here $\Delta t$ is the running time, $N_{{\rm Ge}}$ is the number of
Ge nuclei, $(T_{i},\ T_{i}+\Delta T)$ is the range of recoil energy
in each bin, $\Phi(E_{\nu})$ is the reactor neutrino flux, and
$\theta\left(T_{{\rm max}}(E_{\nu})-T\right)$
is the Heaviside theta function, equal to $0$ for $T_{{\rm max}}(E_{\nu})-T<0$
and 1 for $T_{{\rm max}}(E_{\nu})-T>0$. It is necessary to insert
the $\theta$ function in~(\ref{eq:s-28}) because the cross section
$\frac{d\sigma}{dT}\left(T,\thinspace E_{\nu}\right)$ does not automatically
vanish when $T_{{\rm max}}(E_{\nu})<T$. We take $\Delta t=1$ year,
$\Delta T=0.05$ keV and $N_{{\rm Ge}}=3.32\times10^{25}$, corresponding
to 4 kg natural Ge (with average atomic number $A=72.6$). For $\Phi(E_{\nu})$,
we use a recent theoretical calculation of the flux \cite{Kopeikin:2012zz},
and normalize it to meet the total antineutrino flux
($2.5\times10^{13}\, {\rm s}^{-1}\, {\rm cm}^{-2}$) in CONUS.

\begin{figure}
\centering
\includegraphics[width=8cm]{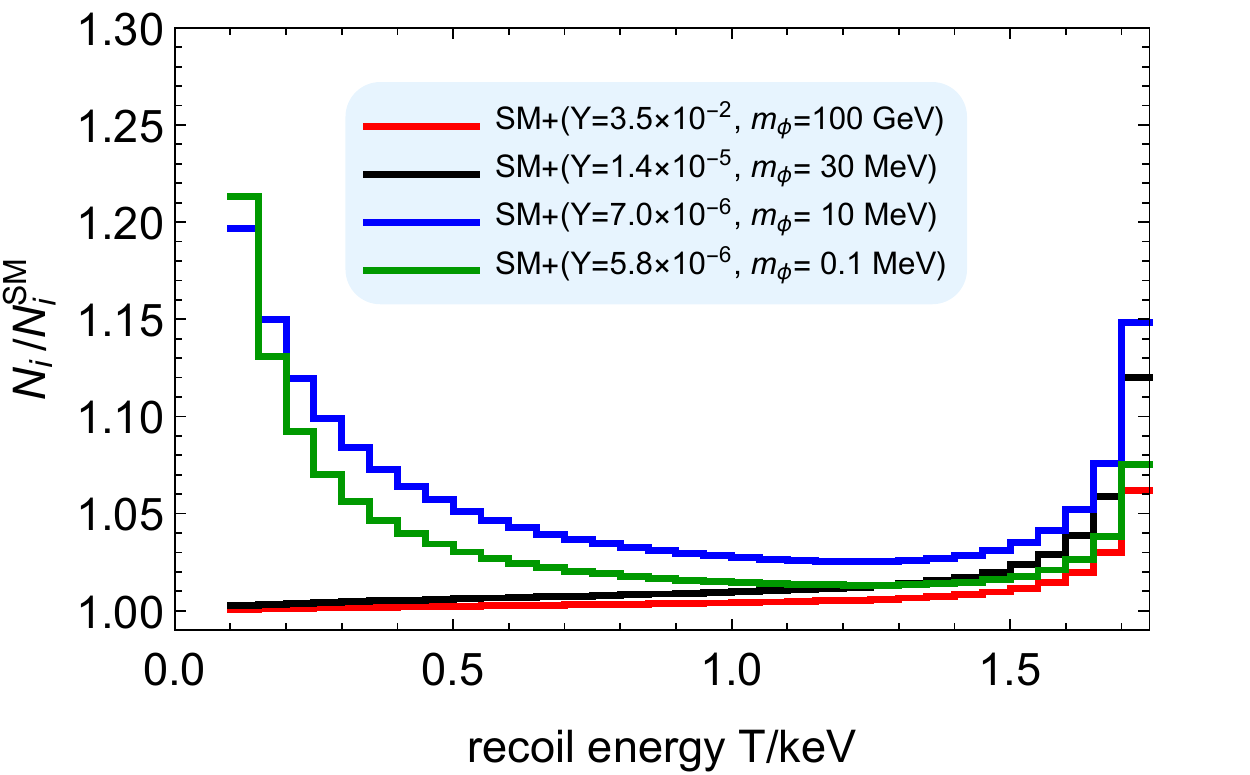}\ \includegraphics[width=8cm]{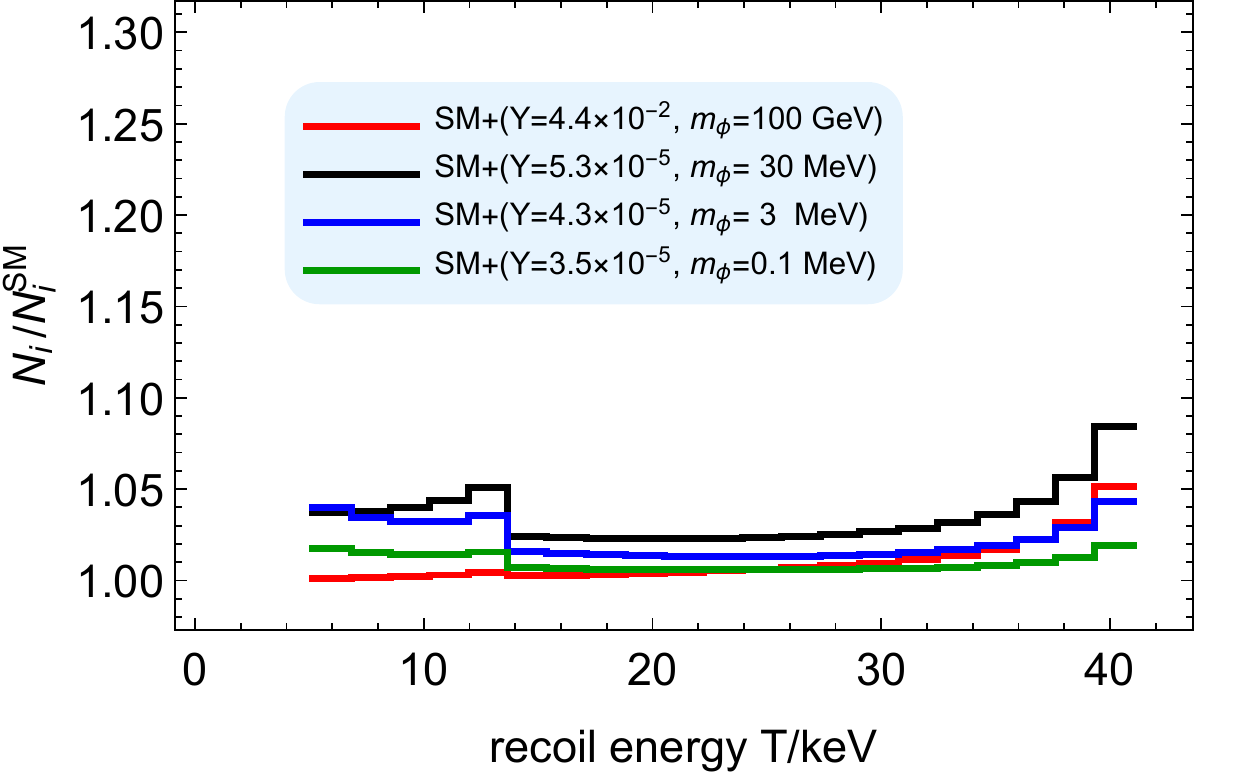}

\caption{\label{fig:ratios}Event excess caused by light scalar bosons in CONUS
(left) and COHERENT (right). The effective coupling $Y$ is defined in  Eq.~(\ref{eq:Y})
and its dependence on the fundamental quark couplings is given by Eq.~(\ref{eq:s-27}). }
\end{figure}

Using Eq.~(\ref{eq:s-28}), we compute the event numbers for several
examples ($m_{\phi}=0.1$ MeV, $10$ MeV, $30$ MeV and $100$ GeV) and
compare them with the SM value in Fig.~\ref{fig:ratios}. The signal strength
is quantified by the ratio $N/N_{0}$ where $N_{0}$ is the SM expectation,
and $N$ contains the additional contributions of light scalar
bosons.
One can see from the figure that the shape of the spectrum when we
include the scalar contribution can be dramatically different, in
particular for low values of $m_\phi$.

To study the sensitivity of CONUS on light scalar bosons, we adopt
the following $\chi^{2}$-function  \cite{Lindner:2016wff},
\begin{equation}
\chi^{2}=\sum_{i}\frac{[(1+a)N_{i}-N_{i}^{0}]^{2}}{\sigma_{{\rm stat},i}^{2}+\sigma_{{\rm sys},i}^{2}}+\frac{a^{2}}{\sigma_{a}^{2}},\label{eq:s-33}
\end{equation}
with
\begin{equation}
\sigma_{{\rm stat},i}=\sqrt{N_{i}+N_{{\rm bkg},\thinspace i}},\thinspace\sigma_{{\rm sys},i}=\sigma_{f}(N_{i}+N_{{\rm bkg},\thinspace i}).\label{eq:s-34}
\end{equation}
The pull parameter $a$ with an uncertainty of $\sigma_a=2 \%$ takes care of the uncertainty in the normalization originating from various sources
 such as the variation of nuclear fuel supply or
the uncertainty of the fiducial mass and distance. Other systematic uncertainties
  that may change the shape of the event spectrum are  parameterized
by $\sigma_{f}$ in Eq.~(\ref{eq:s-34}). Here we assume they are
proportional to the event numbers and take $\sigma_{f}=1\%$. We also
introduce a background in our calculation by adding $N_{{\rm bkg},\thinspace i}$
to the event number in each bin. The background in CONUS is about
1 ${\rm count}/({\rm day}\cdot{\rm keV}\cdot{\rm kg})$. The threshold
of ionization energy detection in CONUS is 0.3 keV, which if divided
by the quenching factor ($\approx0.25$) corresponds to 1.2 keV recoil
energy.  The reactor neutrino flux at $E_{\nu}>8$ MeV has negligible
contributions and also large uncertainties, so we set a cut of $E_\nu$
at 8 MeV, which corresponds to about 1.75 keV recoil energy according
to Eq.~(\ref{eq:s-10}). As a result, in Eq.~(\ref{eq:s-33}) we only
sum over the bins from 1.2 keV to 1.75 keV.

The result is shown in Fig.~\ref{fig:LNC} and Fig.~\ref{fig:LNV}
for the LNC and LNV cases respectively. Although the constraints of
CE$\nu N$S experiments are independent of the LNC/LNV cases, the other
constraints depend on the nature of the interaction, as explained in
Section \ref{sec:limits}. We therefore present the two cases separately.

We also study future improved sensitivities of CONUS by assuming a
100 kg Germanium detector as the target and an improved threshold down to $0.1$ keV.
We assume the corresponding systematic uncertainties are also reduced to a matching
level, $(\sigma_{a},\thinspace\sigma_{f})=(0.5\%,\thinspace0.1\%)$.
This will be possible if the reactor neutrino flux is better understood
due to improved theoretical models and measurements.
The forecast for CONUS100 with 5 years of data taking is also shown in
Figs.\ \ref{fig:LNC}, \ref{fig:LNV} with dashed blue lines.

As mentioned above, the shape information for light scalar masses is
noteworthy in the spectrum, see Eq.\ (\ref{eq:s-9}). It can in fact be
used to determine the value of the mass.
In Fig.\ \ref{fig:signal} we show the potential of CONUS100 for
determining the mass and coupling of the
$\phi$ particle assuming two characteristic examples. As long as the
mass of the scalar is not much larger than the neutrino energy or the
typical momentum exchange, $m_\phi^2 \sim M T \sim  E_\nu^2$,
reconstruction of the mass is possible.

\subsection{COHERENT}

The COHERENT experiment uses a CsI scintillator to detect neutrinos
produced by $\pi^{+}$ and $\mu^+$ decay at rest. In its recent
groundbreaking publication \cite{Akimov:2017ade}
a 6.7$\sigma$ observation of the SM coherent scattering was
announced.
 There are three types of neutrinos in the neutrino flux, $\nu_{\mu}$,
$\overline{\nu}_{\mu}$, and $\nu_{e}$. The first is produced in
the decay $\pi^{+}\rightarrow\mu^{+}+\nu_{\mu}$ while the second
and the third are produced in the subsequent decay $\mu^{+}\rightarrow e^{+}+\overline{\nu}_{\mu}+\nu_{e}$.
Because the first decay is a two-body decay and the pion is at rest,
the produced neutrinos will be monochromatic with energy:
\[
E_{\nu0}=\frac{m_{\pi}^{2}-m_{\mu}^{2}}{2m_{\pi}}\approx29.8\ {\rm MeV},
\]
where $m_{\mu}=105.66$ MeV and $m_{\pi}=139.57$ MeV are the muon
and pion masses, respectively.  Remembering that the muon also decays at rest, the neutrino fluxes are given by \cite{Coloma:2017egw}:
\begin{equation}
\phi_{\nu_{\mu}}(E_{\nu})=\phi_{0}\delta\left(E_{\nu}-E_{\nu0}\right),\label{eq:s-30}
\end{equation}
\begin{equation}
\phi_{\overline{\nu}_{\mu}}(E_{\nu})=\phi_{0}\frac{64E_{\nu}^{2}}{m_{\mu}^{3}}\left(\frac{3}{4}-\frac{E_{\nu}}{m_{\mu}}\right),\label{eq:s-31}
\end{equation}
\begin{equation}
\phi_{\nu_{e}}=\phi_{0}\frac{192E_{\nu}^{2}}{m_{\mu}^{3}}\left(\frac{1}{2}-\frac{E_{\nu}}{m_{\mu}}\right),\label{eq:s-32}
\end{equation}
where $E_{\nu}$ should be in the range $(0,\thinspace m_{\mu}/2)$.
The event numbers are computed by
\begin{eqnarray}
N_{i} & = & \Delta tN_{{\rm Cs/I}}\int_{T_{i}}^{T_{i}+\Delta T}dT\left[\phi_{0}\theta\left(T,\thinspace T_{{\rm max}}(E_{\nu0})\right)\frac{d\sigma}{dT}\left(T,\thinspace E_{\nu0}\right)\right.\nonumber \\
 &  & \left.+\int_{0}^{m_{\mu}/2}dE_{\nu}\left(\phi_{\overline{\nu}_{\mu}}(E_{\nu})+\phi_{\nu_{e}}(E_{\nu})\right)\theta\left(T,\thinspace T_{{\rm max}}(E_{\nu})\right)\frac{d\sigma}{dT}\left(T,\thinspace E_{\nu}\right)\right],\label{eq:s-29}
\end{eqnarray}
which is similar to Eq.~(\ref{eq:s-28}) except that (i) $N_{{\rm Ge}}$
is replaced with $N_{{\rm Cs/I}}$; (ii) for the $\nu_{\mu}$ flux,
the delta function in Eq.~(\ref{eq:s-30}) has been integrated out.
For simplicity, we assume that Cs and I have approximately the same
proton and neutron numbers, $(Z,\thinspace A)=(54,\thinspace130)$.
We also assume that the couplings of neutrinos are flavor universal
($(y_{\nu} )_{\alpha\beta} = y_{\nu} $),
 so we take equal cross section for all neutrino flavors. Using
Eq.~(\ref{eq:s-29}), we also compute the ratio $N/N_{0}$ in COHERENT,
shown in the right panel of Fig.~\ref{fig:ratios}.

\begin{figure}
\centering

\includegraphics[width=14cm]{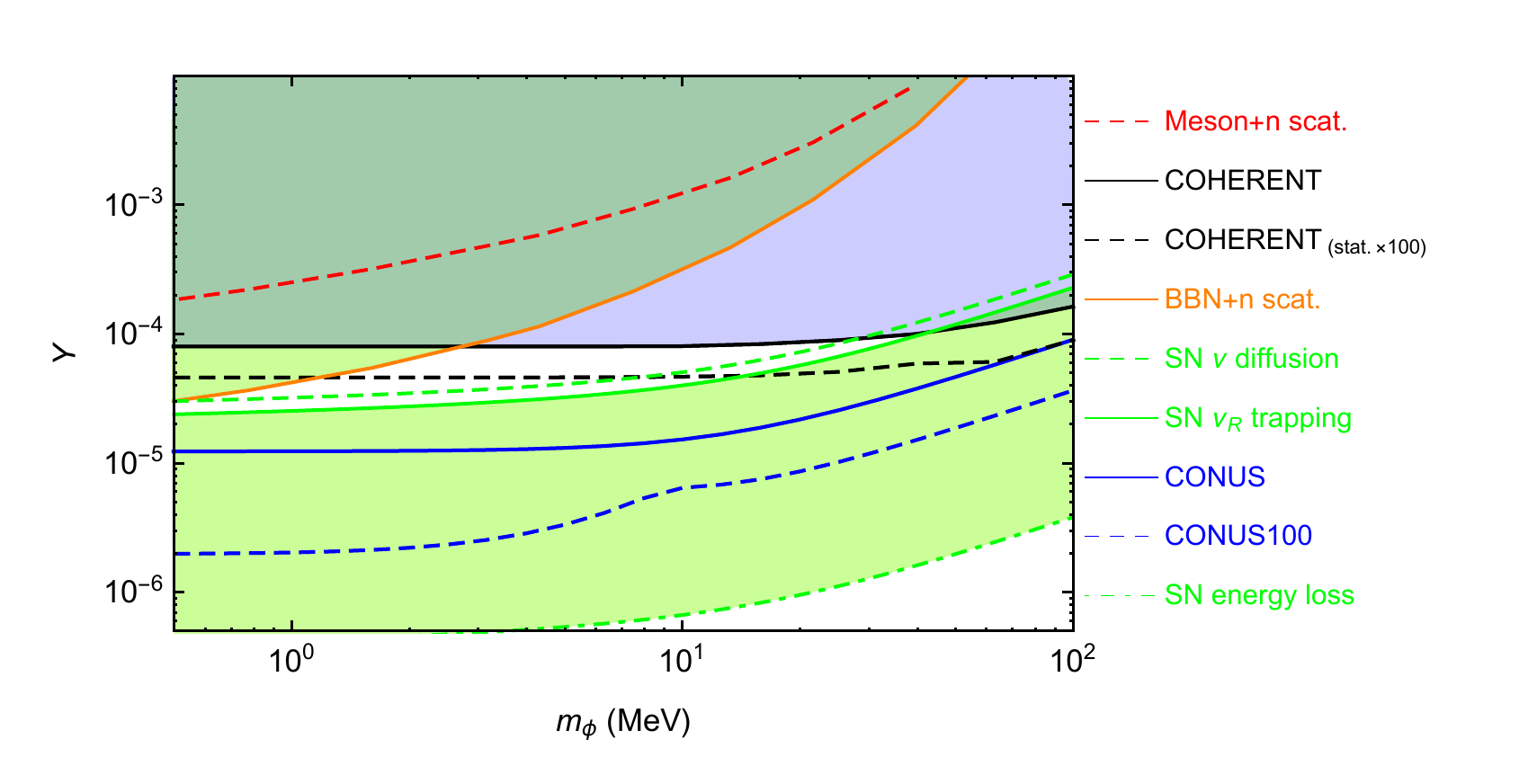}

\caption{Constraints from CE$\nu N$S experiments on $(Y,\thinspace m_{\phi})$
for a lepton number conserving interaction, see Eq.\ (\ref{LNC-Eq}). The black, dashed black,
blue, and dashed blue curves correspond to the 95\% C.L.\ constraint
of the recent COHERENT data, the sensitivities of future COHERENT,
CONUS 4 kg$\times1$ year, and CONUS 100 kg$\times5$ years (light-blue)
respectively.  Various other limits from particle and astroparticle
physics are explained in Section \ref{sec:limits}. \label{fig:LNC}}
\end{figure}

\begin{figure}
\centering

\includegraphics[width=14cm]{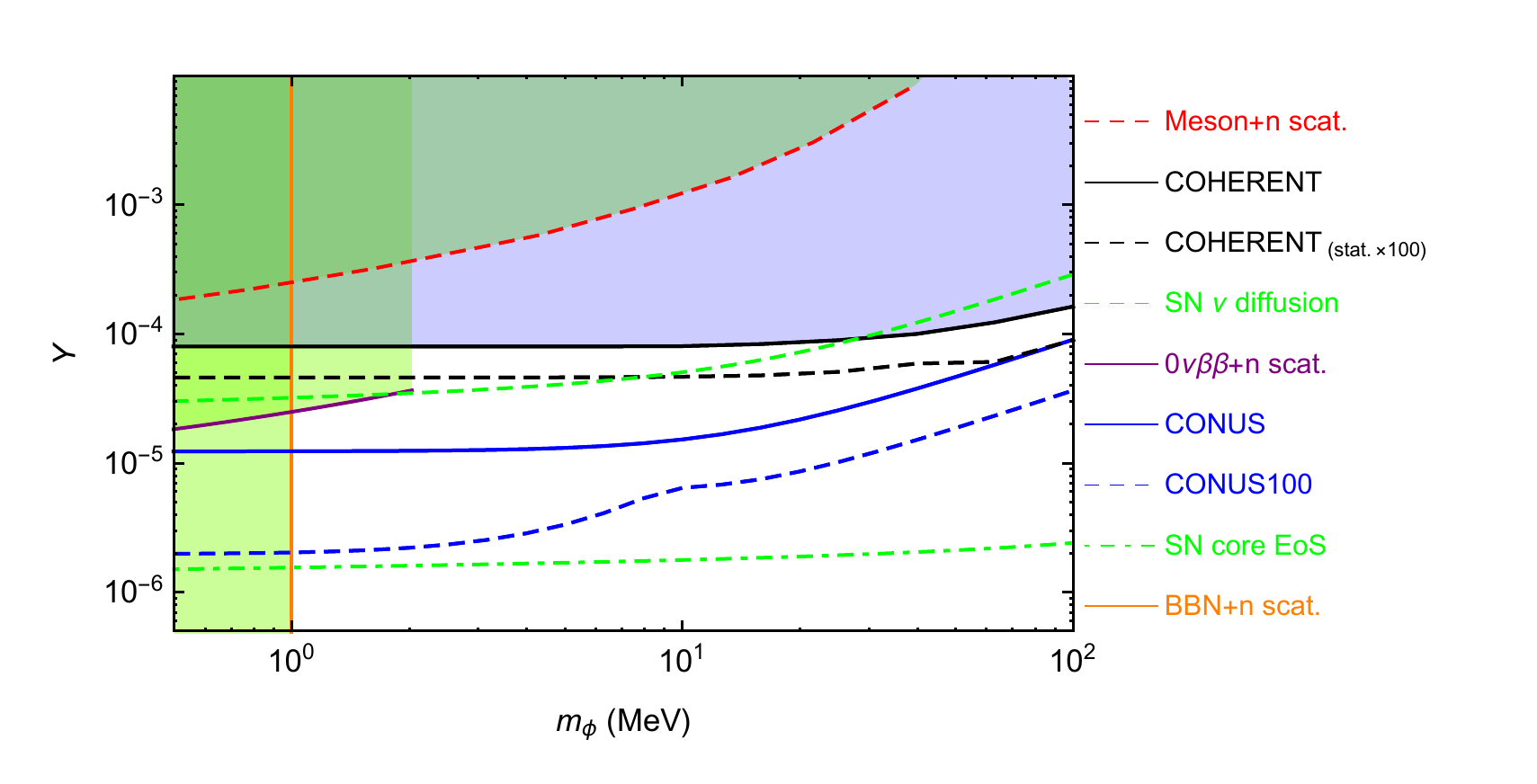}

\caption{Similar to Fig.~\ref{fig:LNC} but for lepton number violating
  interactions, see Eq.\ (\ref{LNV-Eq}) .\label{fig:LNV}}
\end{figure}

In the COHERENT experiment, the recoil energy of the nucleus is converted
to multiple photoelectrons and eventually detected by PMTs. The number
of photoelectrons $n_{{\rm PE}}$ is approximately proportional to
the recoil energy \cite{Akimov:2017ade}:
\begin{equation}
n_{{\rm PE}}\approx1.17\frac{T}{{\rm keV}}.\label{eq:s-35}
\end{equation}
For $n_{{\rm PE}}>20$, the signal acceptance fraction is about 70\%
(cf.\ Fig.~S9 of \cite{Akimov:2017ade}). This number drops down quickly
for smaller $n_{{\rm PE}}$, and becomes approximately zero for $n_{{\rm PE}}<5$.
This implies that the threshold for $T$ is about 4 keV in COHERENT.
Using Eqs.~(\ref{eq:s-29}, \ref{eq:s-35}) and the signal acceptance
fraction data, we can study the constraint of the COHERENT data (from
Fig.~3 of \cite{Akimov:2017ade}) on light scalar bosons. The
SM expectation is also provided by \cite{Akimov:2017ade}  which can
be used to compute the total normalization factor. We directly use
the relevant uncertainties provided by \cite{Akimov:2017ade}.  The result is shown
in Figs.~\ref{fig:LNC} and \ref{fig:LNV} as well. Since reactor
neutrinos provide much larger event numbers, CONUS limits will be
better, though of course limited to the electron-type couplings,
whereas COHERENT will also have muon-type neutrinos.

\begin{figure}

\centering
\includegraphics[width=\textwidth]{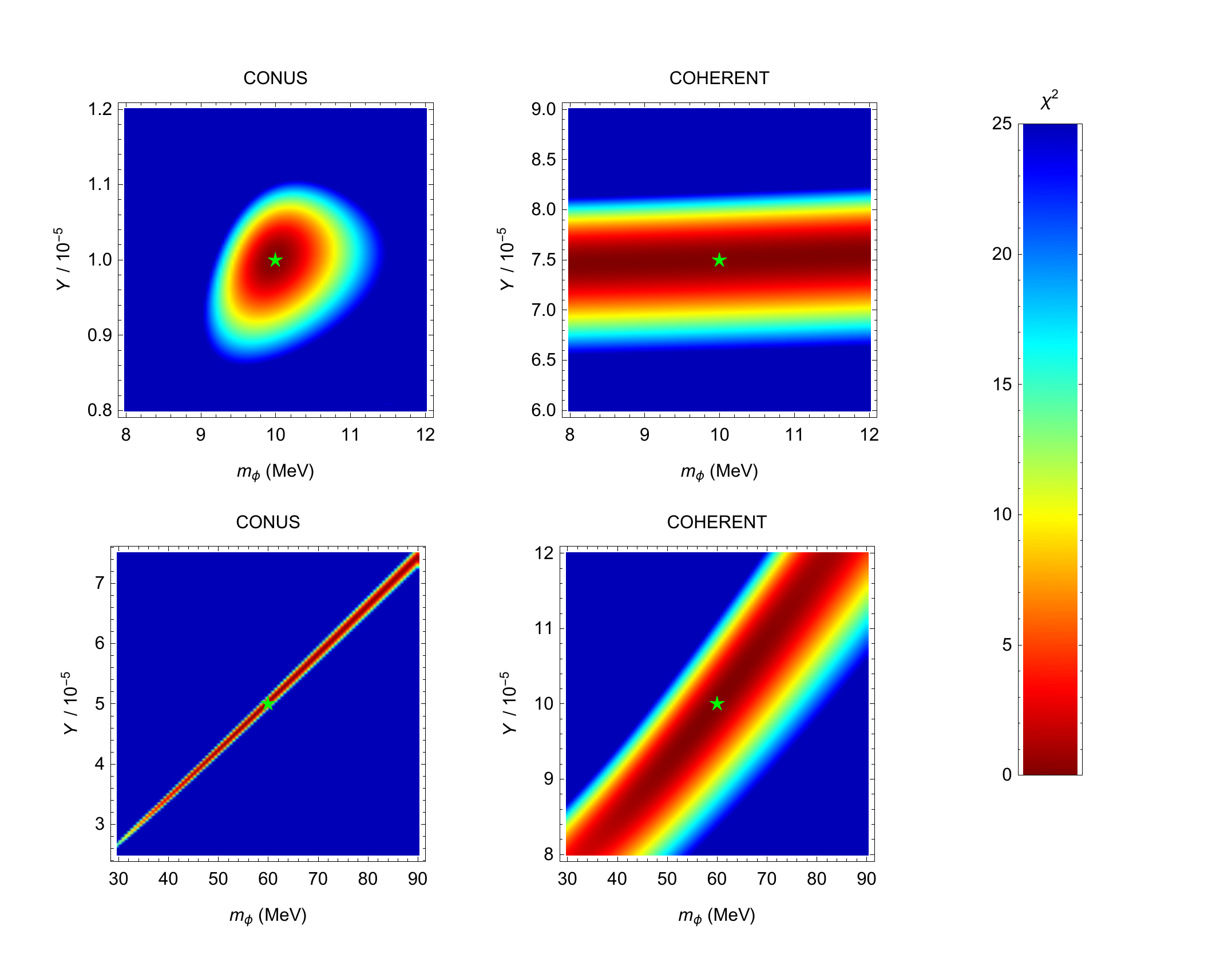}

\caption{\label{fig:signal}Measurements of the mass $m_{\phi}$ and coupling
$Y$ in CONUS100 (left panels) and COHERENT (stat.\ $\times$ 100) (right panels) assuming the presence of a scalar boson, with the true values indicated by the green stars.
}

\end{figure}

In the future, the COHERENT experiment will further develop the detection
of  CE$\nu N$S with different targets\footnote{See: http://webhome.phy.duke.edu/\textasciitilde{}schol/COHERENT\_Yue.pdf},
including 30 kg liquid argon, 10 kg high purity Ge, and 185 kg NaI
crystal. A complete study on the future sensitivities of future COHERENT
including all the different targets and different detection technology
is beyond  the scope of this paper. Considering that the total fiducial
mass compared to the current value (14.6 kg CsI) will be increased
by a factor of $\sim20$, plus a prolonged running for few years,
at best the statistics may be increased by a factor of $100$, which
corresponds to a reduction of the statistical uncertainties by a
factor of $10$.
It is therefore reasonable to assume that the uncertainties of future
measurement will be reduced by a factor between 1 and 10. To show
the sensitivity of future COHERENT versions on the light scalar coupling, we
plot a black dashed curve in Fig.~\ref{fig:LNC} and Fig.~\ref{fig:LNV}
assuming the uncertainties (both systematical and statistical) are
reduced by a factor of 10. For the sake of definiteness,
we take the liberty to denote this potential situation as COHERENT (stat.\ $\times$ 100).
Again, the determination of the mass of the
scalar particle is possible if its mass lies below the typical
neutrino energy. As seen in Fig.\ \ref{fig:ratios}, the spectral
distortion due to the scalar exchange is less dramatic as for
CONUS. This is mostly caused by the larger energy of the neutrinos, the
momentum exchange and nuclear recoil. 
 Fig.\ \ref{fig:signal} shows the potential of the assumed
 future COHERENT version for determining the mass and coupling of the 
 $\phi$ particle, assuming two characteristic examples. Due to the
 larger energy of COHERENT, and also because of the smaller
 statistics, the reconstruction potential is less promising
 compared to experiments based on reactor neutrinos.

\section{Interpretation of the Results \label{sec:disc}}
Figs.\ \ref{fig:LNC} and  \ref{fig:LNV}   show the
constraints and limits on the relevant combination of $Y$ couplings
versus the mass of the scalar for  lepton number
conserving and lepton number violating interactions, respectively. To draw these
lines the coupling of $\phi$ to neutrinos is taken to be flavor
universal.  Each limit is however sensitive to a different flavor
structure. Let us start by discussing the bounds which apply for both
lepton number violating and lepton number conserving interactions. The
red-dashed lines show the constraint on $Y \simeq \sqrt{y_{\nu}  C_n}$
from combining the upper bounds on $C_n$ and $y_{\nu} $ from the $n$-Pb
scattering and meson decay experiments. As seen from the figures, this
bound is relatively weak. The present bound from COHERENT shown by a
solid black line is already well below this combined bound. The bounds
from meson decay are sensitive to $\sum_\alpha (y_{\nu} )_{e \alpha}^2$
and $\sum_\alpha (y_{\nu} )_{\mu \alpha}^2$. Since the beam at the
COHERENT experiment is composed of
$\nu_\mu$ $\bar \nu_\mu$ and $\nu_e$ fluxes, it will be sensitive to similar flavor composition.
Our forecast for the future bound by the COHERENT experiment (a factor
10 smaller uncertainties) 
is shown by dashed black line; the bound
on $Y$ can be improved by a factor of 2. The blue solid and dashed lines are the upper bounds that CONUS can set with $1~{\rm year} \times 4~{\rm kg}$ and $5~{\rm year} \times 100~{\rm kg}$ of data taking, respectively.
 As seen from these figures, CONUS can improve the bound by one or two orders of magnitudes. Since CONUS is a reactor neutrino experiment, it can only probe $ C_n^{1/2}(\sum_\alpha |y_{\nu} |_{e \alpha}^2)^{1/4}$.
 Since the uncertainties of CONUS are mainly limited by statistics,
 the bound that it can set on the cross section $\sigma_\phi$ scales as $t^{-1/2}$ with data taking time.
 Since $\sigma_\phi\propto Y^4$, the bound on $Y$ will scale as $t^{-1/8}$.

The violet curve in Fig.\  \ref{fig:LNV} up to 2.4 MeV is the combined bound from the $n$-Pb
scattering on $C_n$ and from double beta decay on $(y_{\nu} )_{ee}$.  CONUS with only one year of data taking
can provide a stronger bound.
 The dashed green lines in Figs.\ \ref{fig:LNC} and  \ref{fig:LNV} denoted
``SN $\nu$ diffusion'' show the limits resulting essentially from a neutrino-nucleon
scattering cross section due to $\phi$ exchange being equal to that in the SM.
As we discussed before, in the vicinity of this line supernova evolution and emitted neutrino flux will be dramatically
affected. As seen from the figures, the CONUS experiment with 1 year of data taking
can already probe all this range.  The green area between solid and dotted-dashed green lines in Fig.\ \ref{fig:LNV} is
ruled out by supernova cooling and $\nu_R$ trapping considerations.

In the LNC case, the orange line in Fig.\ \ref{fig:LNC} denoted by ``BBN + n scat.'' shows the combined bound from $n$-Pb
scattering and BBN. As seen from the figure, for $m_\phi > 3$ MeV, the bound from COHERENT is already stronger.
Both for LNC and LNV cases, a $\phi$ particle with $m_\phi \in (1.5-3)$ MeV and $y_{\nu}  \sim {\rm few} \times 10^{-5}$ can significantly affect BBN.
 As seen from the figures this mass range can be probed by CE$\nu$NS experiments. For
 $\sqrt{C_n |y_{\nu} |_{e\alpha}}$ above the dotted-dashed green line in Fig.\ \ref{fig:LNV}, the equation of state in supernova inner core will drastically change because of  $\nu_e +n \to \bar{\nu}_\alpha +n$ scattering. As seen from the figure, a significant part of the parameter space above this line can be probed by CONUS.
We can therefore deduce that coherent scattering results may have dramatic impact on SN and BBN physics.


Fig.~\ref{fig:signal} displays the prospect of measuring $m_\phi$
and $Y$ by our future versions of CONUS and COHERENT, assuming
true values of $(m_{\phi},\ Y)$ are $(10\ {\rm MeV},\ 10^{-5})$
or $(60\ {\rm MeV},\ 5\times10^{-5})$ for CONUS100, and $(10\ {\rm MeV},\ 7.5\times10^{-5})$
or $(60\ {\rm MeV},\ 10^{-4})$ for COHERENT (stat.\ $\times$ 100). Here for comparison,
we choose the same masses for the two experiments. However, the couplings
in COHERENT (stat.\ $\times$ 100) are set to larger values to lead to similar precision as
CONUS100 (cf.\ Figs.~\ref{fig:LNC} and
\ref{fig:LNV}). Even with larger couplings, COHERENT still cannot
measure $(m_{\phi},\ Y)$ as good as CONUS100. As shown in Fig.~\ref{fig:signal},
for true values $m_{\phi}=10$ MeV and $Y=10^{-5}$, the mass and
coupling can be determined with better than 10\% accuracy by CONUS100.
In comparison, COHERENT (stat.\ $\times$ 100) loses its capability to determine $m_{\phi}$
but it has still reasonable precision in determining $Y$ provided
that $Y$ is large enough (close to its present bound). This is understandable
because $m_{\phi}=$ 10 MeV is larger than the typical energy-momentum
transfer in the CONUS100 ($m_{\phi}^{2}\gs MT\sim E_{\nu}^{2}$)
but is smaller than the energy-momentum transfer in COHERENT
($m_{\phi}^{2}\ls MT\sim E_{\nu}^{2}$). Another reason is the better
statistics in CONUS. 
If the mass is raised to 60 MeV, then both lose their ability to determine
the mass and the coupling separately. They however maintain  their sensitivity
to $Y/m_{\phi}$.


 Discovering a positive signal for the effects of $\phi$ by CONUS will have drastic consequences
for the analysis of supernova evolution. If a value of $Y$ below the
green solid line in Fig.\   \ref{fig:LNC} is found, the
supernova cooling bounds tell us that interaction cannot involve
$\nu_R$ so the interaction should be of lepton number violating form.
  If $\phi$ turns out to have a mass around 2 MeV, it will be more intriguing as it may be discovered at double beta experiments. If, however, double beta decay searches fail to discover $\phi$ with expected mass and coupling, we may
draw a conclusion that $(y_{\nu} )_{ee}\ll  (y_{\nu} )_{e\mu},
(y_{\nu} )_{e\tau}$. In any case, in analyzing BBN, effects of such light $m_\phi$ with sizeable $Y$ has to be taken into account.
Comparing Figs.\  \ref{fig:LNC} and  \ref{fig:LNV}, we conclude that
because of the BBN bounds, discovery of $m_{\phi}<1 $ MeV will indicate
LNC interaction with light right-handed neutrinos with immediate
consequences for supernova evolution.

If CONUS and/or COHERENT finds $Y\sim 10^{-5}$, the chances of finding a signal for $\phi$ in meson
decay experiments as well as in $n$-Pb scattering experiments increase.
If CONUS finds a signal for $Y>5 \times 10^{-5}$, this means that signals for $K^+ \to e^+ \nu \phi$
and for  $\pi^+\to e^+ \nu \phi$ will be within reach of next
generation \cite{Bakhti:2017jhm}, and new $n$-Pb measurements would be
very interesting.
If CONUS finds $Y<10^{-5}$ and if meson decay
experiments find $y_{\nu}  \gs 10^{-4}$, we would conclude
$C_n<10^{-6}$, making it difficult to see an effect on $n$-Pb
scattering experiments. Similarly $Y<10^{-5}$ and $C_n \sim 10^{-4}$
(close to the present bound) would imply $y_{\nu}  <10^{-6}$.

We conclude this section by mentioning some possibilities on inferring the flavor structure or type of interaction
that arise due to the complementarity of the various sources and
limits.
If the  $\nu_e$ couplings are large enough to be within the reach of
COHERENT (that is if $Y>5 \times 10^{-5}$), the CONUS experiment will easily determine
$\sum_\alpha |(y_{\nu} )_{e\alpha}|^2 C_N^2$, where $\alpha$  runs over all active flavors for the LNV case (over all
light right-handed species for LNC case).  The information by COHERENT
can then determine $\sum_\alpha |(y_{\nu} )_{\mu \alpha}|^2C_N^2$.
If COHERENT alone would be able to distinguish the flavor content of
the events (e.g.\ by timing cuts), information on flavor structure of $y_{\nu} $
(i.e.\ on $\sum_\alpha |(y_{\nu} )_{\mu \alpha}|^2/ \sum_\alpha
|(y_{\nu} )_{e \alpha}|^2$) could be extracted.
In the special case that $|(y_{\nu} )_{e\alpha}|\ll |(y_{\nu} )_{\mu\alpha}|$, it may be possible that
COHERENT will discover new effect but CONUS will report null results
for new physics discovery.

\section{Summary and concluding remarks\label{sec:concl}}

Coherent elastic neutrino-nucleus scattering can probe both new light as well
as heavy physics. Focussing here on the light case we demonstrated the
discovery potential of current and future coherent scattering
experiments  on the mass and coupling of scalar particles interacting
with neutrinos and quarks. The shape of the nuclear recoil spectrum is
distorted by the scalar interaction, and allows even to determine the
mass of the scalar, if its mass is around the energy of the scattered
neutrinos. Even current limits by COHERENT are competitive with a combination of bounds from BBN and from various terrestrial experiments such as meson decay and neutron-scattering experiments.
 Moreover, these bounds probe areas in parameter space
that can have important consequences for BBN and supernova evolution, in
particular for lepton number violating interactions.
Future versions of the experiment or upcoming
reactor experiments such as CONUS will reach not yet explored areas in
parameter space.

\begin{acknowledgments}
We thank Giorgio Arcadi, Tommy Ohlsson, Kate Scholberg and Stefan
Vogl  
for many helpful discussions.
YF 
thanks MPIK Heidelberg where a part of this work was done for their hospitality.
 This project has received funding from the European Union\'~\!s Horizon 2020 research and innovation programme under the Marie Sklodowska-Curie grant agreement No 674896 and No 690575.
YF is also grateful to ICTP associate office for partial financial support.
WR is supported by the DFG with grant RO 2516/6-1 in the Heisenberg program.
\end{acknowledgments}

\appendix

\section{Suggestions for underlying electroweak symmetric models\label{model-EW}}
In this section we show how one can build a toy model symmetric under $SU(2) \times U(1)$ that can give rise to effective
coupling of form
$$\bar{q}\Gamma_q q \phi\equiv \bar{q} (C_q+i\gamma^5 D_q) q \phi$$ as
well as the ones shown in Eqs.\ (\ref{LNC-Eq}, \ref{LNV-Eq}). The coupling of $\phi$ to nuclei should of course arise from its coupling  to quarks.  The latter  can originate from the mixing of the singlet scalar $\phi$
with an electroweak doublet, $\Phi$. Taking  complex couplings of
form \begin{equation} Y_u \bar{u}_R \Phi^T C Q+{\rm H.c.} \ {\rm and}
  \ Y_d \bar{d}_R \Phi^\dagger Q+{\rm H.c.} \end{equation} and a
mixing of $\beta$ between $\phi$ and the neutral component of $\Phi$,
we find the coupling of $\phi$ to the $u$ and $d$ quarks respectively
to be given by $\Gamma_{u }=\sin \beta \, Y_u^*$ and  $\Gamma_{d
}=\sin \beta Y_d^*$, or equivalently
\begin{equation} C_q={\cal R}[Y_q] \sin \beta  \ \ \ {\rm and} \ \ \
  D_q={\cal I}[Y_q] \sin \beta .\end{equation}  Notice that taking $Y_q$ to be real, the coupling will be parity invariant and therefore $D_q=0$.

The most economic solution is to identify $\Phi$ with the SM Higgs.
Remember that the couplings of the SM Higgs to quarks of first
generation are ${\cal O}(10^{-5})$. Moreover, the mixing of $\phi$
with SM Higgs cannot exceed ${\cal O}(10^{-2})$, 
otherwise the rate of invisible decay mode $Br(H \to \phi \phi)$ will
exceed the experimental limits. Combining these, we conclude that in
case $\phi$ is taken to be  the SM Higgs, $|\Gamma_{q\phi}|\sim 10^{-2} \,
m_q/\langle H^0\rangle$. As we have seen in Sec.\
\ref{Ff}, the contributions from quarks of different generations to
the coupling of a nucleus to $\phi$ will be of the same order and too
small to lead to discernable effects on current CE$\nu$NS experiments.

Taking thus $\Phi$ to be a new doublet, its coupling to quarks can in
principle be as large as ${\cal O}(1)$.
Taking $$V(\phi,\Phi)=\frac{m_1^2}{2} \phi^2+m_2^2 |\Phi|^2+ (A \phi H^\dagger \cdot \Phi+{\rm H.c.}),$$
we obtain $\sin \beta= A \langle H\rangle /(m_2^2-m_1^2)$,
$m_\Phi\simeq m_2$ and $m_\phi\simeq \sqrt{m_1^2-m_\Phi^2 \sin^2
  \beta}$. To avoid a need for fine-tuned cancelation $m_\Phi \sin
\beta$ should be smaller than ${\cal O}(m_\phi)$. For $m_\phi \sim 5$
MeV and $m_\Phi \sim 1$ TeV, this implies $\sin \beta \sim
10^{-5}$. For $\Gamma_{u} \sim \Gamma_{d} \sim 10^{-5}$, naturalness
(i.e., absence of fine-tuned cancelation) requires the mass of $\Phi$
to be within the reach of the LHC and its coupling to $u$ and $d$
quarks to be ${\cal O}(1)$ which in turn promises a rich phenomenology
at the LHC.

By turning to the LNC interaction $\bar{\nu}_R \nu_L$, a  mechanism
similar to the one described above can provide a lepton number
conserving interaction of $\phi \bar{\nu}_R \nu_L$ through mixing of
$\phi$ with neutral component of the $\Phi$ doublet. For a lepton
number violating coupling $\phi \nu^T C \nu$, two scenarios can be
realized:
\begin{itemize} \item The $\phi \nu^T C \nu$ coupling can be obtained by the mixing of $\phi$ with a neutral component of an electroweak triplet $\Delta$ which couples to the left-handed lepton doublet as $L^T C \Delta L$. The quantum numbers of $\Delta$ should be the same as the triplet scalar whose tiny vacuum expectation value is responsible for neutrino mass in the type II seesaw mechanism. Identifying these two, the flavor structure of the $\phi$ coupling to neutrinos will be determined by that of neutrino mass: $(C_\nu)_{\alpha \beta} , (D_\nu)_{\alpha \beta} \propto (m_\nu)_{\alpha \beta}$. Naturalness (i.e., $m_\phi \sim 1 ~{\rm MeV}\ll m_{\Delta}\sim {\rm TeV}$ without fine tuned cancelation) again implies that $\Delta$ is not much heavier than TeV and its coupling to leptons are of order of 1 which promises rich phenomenology at the LHC such as production of $\Delta^{++}$ and its decay into a same sign pair of charged leptons.
\item Another scenario that can provide coupling of form  $\phi \nu^T
  C \nu$ is suggested in \cite{Blum:2014ewa}. The scenario is very
  similar to the inverse seesaw mechanism for generating mass for
  neutrinos and requires a Dirac fermion singlet $\Psi$ with the following Lagrangian
    \begin{equation} 
{\cal L} = 
M_\Psi \bar{\Psi}\Psi+y \bar{\Psi}_R H^T CL+ y^\prime \phi {\Psi}_R^T  C\Psi_R.\end{equation}
    When $M_\Psi \gg m_\phi$, we can integrate out $\Psi$ and arrive
    at $C_\nu =(y \langle H\rangle /M_\Psi)^2 y^\prime$. As shown in
    \cite{Blum:2014ewa}, $C_\nu$  even as large as  $10^{-3}$ can be
    obtained by this mechanism. Moreover, if $\phi$ develops a vacuum
    expectation value, an inverse seesaw mechanism for neutrino mass
    generation will emerge and the flavor structure of $C_\nu$ and
    neutrino mass matrix will be similar. In this scenario, the SM
    Higgs will have an invisible decay mode $H\to \bar{\Psi} \nu_L$
    governed by  $y^2$.
    \end{itemize}

The discovery of a scalar field in coherent scattering  experiments
will therefore, at least in the models outlined here, hint towards a rich new collider phenomenology.

\section{\label{sec:calc}Calculation of the cross section}

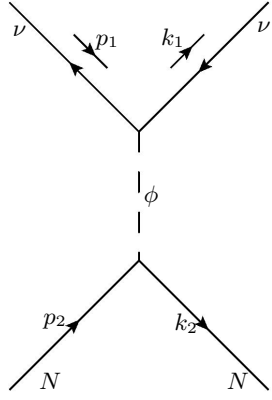
\begin{figure}
\centering

\begin{overpic}[width=3.5cm]{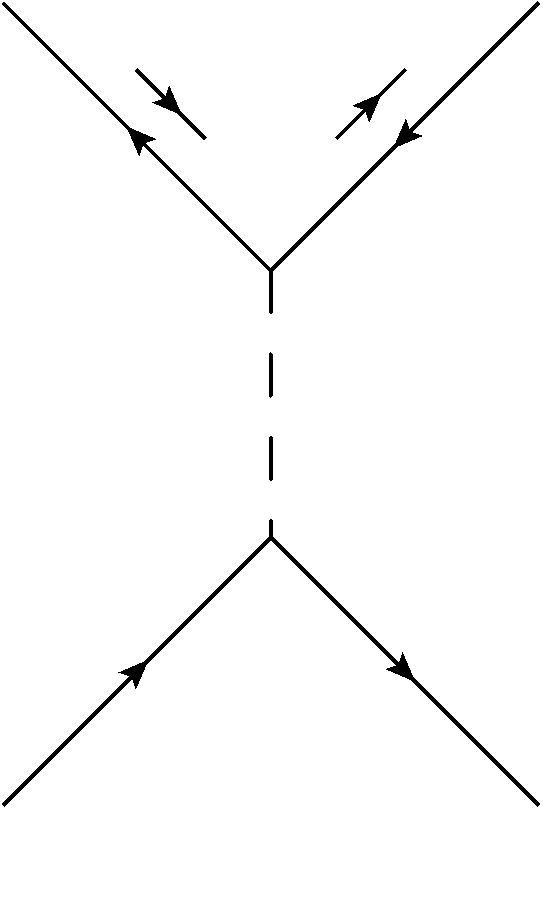}
\put (1,92) {$\nu$} \put (57,93) {$\nu$}
\put (7,11) {$N$} \put (50,11) {$N$}
\put (31,54) {$\phi$}
\put (20,90) {$p_{1}$} \put (35,90) {$k_{1}$}

\put (8,26) {$p_2$} \put (38,24.5) {$k_2$}

\end{overpic}

\caption{\label{fig:Feyn}The Feynman diagram of coherent $\overline{\nu}N$
scattering mediated by the new light scalar boson.}
\end{figure}

In this appendix, we give the analytic calculation of $\overline{\nu}N$
and $\nu N$ cross sections, assuming a lepton number conserving
interaction with $\phi$. The cross section for the lepton number violating case is identical. Let us first focus on the $\overline{\nu}N$
case. The initial and final momenta are denoted in the way shown in
Fig.~\ref{fig:Feyn}. The scattering amplitude of this diagram is
\begin{equation}
i{\cal M}_{\phi}=\overline{v}^{s}(p_{1})P_{R}
\left(i\Gamma_{\nu\phi}\right)
v^{s'}(k_{1})\frac{-i}{q^{2}-m_{\phi}^{2}}\overline{u}^{r'}(k_{2})
\left(i\Gamma_{N\phi}\right)u^{r}(p_{2}),\label{eq:s-3}
\end{equation}
where $\Gamma_{\nu\phi}\equiv C_{\nu}+D_{\nu}i\gamma^{5}$ and
\begin{equation}
q=p_{1}-k_{1},\ q^{2}=-2MT.\label{eq:s-4}
\end{equation}
We have inserted a right-handed projector $P_{R}=(1+\gamma^{5})/2$
in Eq.~(\ref{eq:s-3}) because the initial antineutrino produced by
the charged current interaction should be right-handed. Because of
$P_{R}$, if the initial antineutrino is left-handed, the amplitude
automatically vanishes. We can therefore  sum over all the spins and
apply the trace technology:
\begin{eqnarray}
|i{\cal M}_{\phi}|^{2} & = & \frac{1}{(2MT+m_{\phi}^{2})^{2}}{\rm tr}\left[\gamma\cdot p_{1}P_{R}\Gamma_{\nu\phi}\gamma\cdot k_{1}\Gamma_{\nu\phi}P_{L}\right]\frac{1}{2}{\rm tr}\left[(\gamma\cdot k_{2}+M)\Gamma_{N\phi}(\gamma\cdot p_{2}+M)\Gamma_{N\phi}\right].\label{eq:s-12}\\
 & = & \frac{1}{(2MT+m_{\phi}^{2})^{2}}{\rm tr}\left[\gamma\cdot p_{1}P_{R}\gamma\cdot k_{1}(C_{\nu}^{2}+D_{\nu}^{2})P_{L}\right]2M\left[C_{N}^{2}(2M+T)+D_{N}^{2}T\right].\label{eq:s-12-1}
\end{eqnarray}
For $\nu N$ scattering, one needs to change Eq.~(\ref{eq:s-3}) to
\begin{equation}
i{\cal M}_{\phi}=\overline{u}^{s}(p_{1})P_{L}\left(i\Gamma_{\nu\phi}\right)
u^{s'}(k_{1})\frac{-i}{q^{2}-m_{\phi}^{2}}
\overline{u}^{r'}(k_{2})\left(i\Gamma_{N\phi}\right)u^{r}(p_{2}).\label{eq:s-3-1}
\end{equation}
Consequently, one has to interchange $P_{R}\leftrightarrow P_{L}$
in Eq.~(\ref{eq:s-12}). From Eq.~(\ref{eq:s-12-1}) we can see that
$P_{R}\leftrightarrow P_{L}$ does not change the result so the cross
sections are equal for  $\overline{\nu}N$ and ${\nu}N$ scattering.
From Eq.~(\ref{eq:s-12-1})  we obtain
\begin{equation}
\frac{d\sigma_{\phi}}{dT}=\frac{M|y_{\nu}|^{2}}{4\pi(2MT+m_{\phi}^{2})^{2}}\left[C_{N}^{2}\frac{MT}{E_{\nu}^{2}}+
\left(C_{N}^{2}+D_{N}^{2}\right)\frac{T^{2}}{2E_{\nu}^{2}}\right],
\label{eq:s-9-1}
\end{equation}
which is the cross section for both neutrino and antineutrino scattering.

\bibliographystyle{apsrev4-1}
\bibliography{ref}

\end{document}